\documentclass{article} 	
\usepackage{spconf,amsmath,graphicx,amssymb,booktabs,bm,pifont,multirow}
\usepackage{bbding}
\usepackage{pifont}
\usepackage{wasysym}
\usepackage{textcomp,array,booktabs}
\usepackage[hidelinks]{hyperref}
\newcommand*{\email}[1]{\href{mailto:#1}{\nolinkurl{#1}} }

\makeatletter
\def\endthebibliography{%
	\def\@noitemerr{\@latex@warning{Empty `thebibliography' environment}}%
	\endlist
}
\makeatother

\title{Enhanced Memory Network: The novel network structure for Symbolic Music Generation}
\name{Jin Li$^{1,2}$, Haibin Liu$^{1,3}$, Nan Yan$^{1,2}$, Lan Wang$^{1,2}$}
\address{$^1$CAS Key Laboratory of Human-Machine Intelligence-Synergy Systems, \\
	$^2$Guangdong-Hong Kong-Macao Joint Laboratory of Human-Machine Intelligence-Synergy Systems, \\
	Shenzhen Institute of Advanced Technology, Chinese Academy of Sciences, Shenzhen, China \\
	$^3$Chinese Academy of Sciences, Beijing, China \\
	\email{{li.jin, hb.liu1, nan.yan, lan.wang}@siat.ac.cn}
}

\begin{document}
	%
	\maketitle
	\begin{abstract}
		Symbolic melodies generation is one of the essential tasks for automatic music generation. Recently, models based on neural networks have had a significant influence on generating symbolic melodies. However, the musical context structure is complicated to capture through	deep neural networks. Although long short-term memory (LSTM) is attempted to solve this problem through learning order dependence in the musical sequence, it is not capable of capturing musical context with only one note as input for each time step of LSTM. In this paper, we propose a novel Enhanced Memory Network (EMN) with several recurrent units, named Enhanced Memory Unit (EMU), to explicitly modify the internal architecture of LSTM for containing music beat information and reinforces the memory of the latest musical beat through aggregating beat inside the memory gate. In addition, to increase the diversity of generated musical notes, cosine distance among adjacent time steps of hidden states is considered as part of loss functions to avoid a high similarity score that harms the diversity of generated notes. Objective and subjective evaluation results show that the proposed method achieves state-of-the-art performance. Code and music demo are available at https://github.com/qrqrqrqr/EMU
		
	\end{abstract}
	\section{Introduction}
	In recent years, deep neural networks (DNNs) have obtained significant improvement in varieties of artificial intelligence areas such as image recognition \cite{krizhevsky2012imagenet}, neural language translation \cite{young2018recent}, automatic speech recognition \cite{hinton2012deep}, and automatic music generation \cite{ji2020comprehensive}. Automatic music generation has two mainstream music generation approaches include audio music generation \cite{yang2017midinet, hadjeres2017deepbach, huang2018music} and symbolic music generation \cite{oord2016wavenet, engel2017neural, mehri2016samplernn}. In this paper, we focus on symbolic music generation \cite{wu2019hierarchical}.
	
	\begin{figure*}[]
		\centering
		\includegraphics[width=16cm]{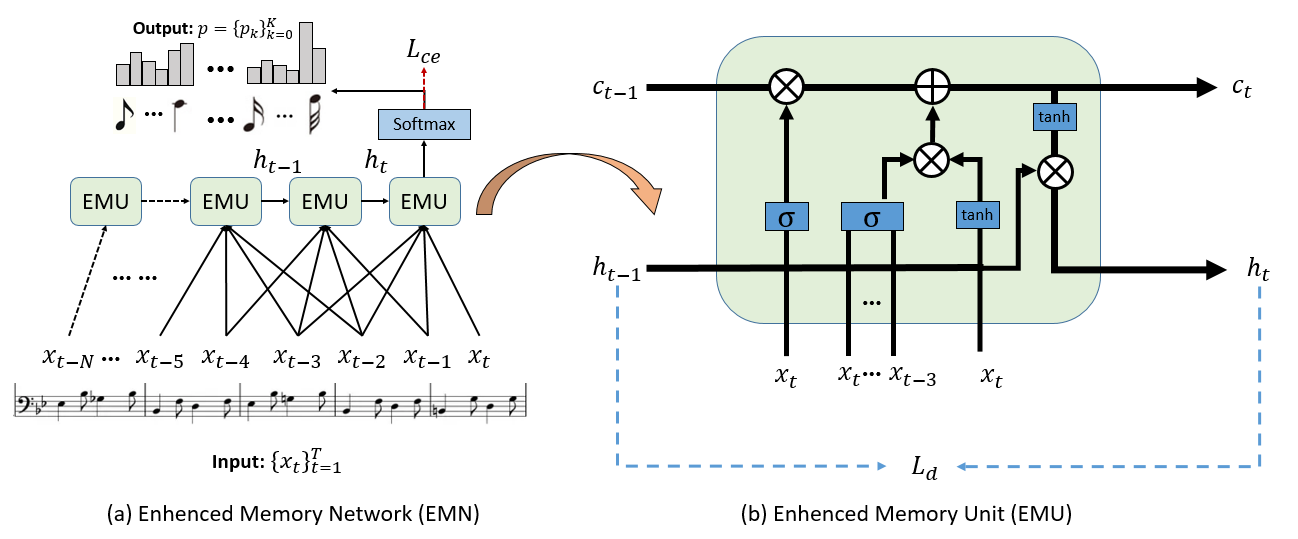}
	\caption{Illustration of our Enhanced Memory Unit and EMN model. (a) Given an input musical encodings $\{x_t\}_{t=1}^T$, EMN models the musical sequence and outputs the probability distribution $p$ of the next musical note over all music notes $K$ and holding state. (b) The EMU extends LSTM with additional inputs, four notes with learnable parameters (i.e. from $x_t$ to $x_{t-3}$) and incorporate cosine distance loss $\mathcal{L}_d$ between two neighboring hidden states to increase the diversity of generated musical notes.}
		\label{fig:architecture}
	\end{figure*}
	
	Symbolic music generation is the process of generating musical notes with control parameters such as instrument type and music style. Many models attempt to incorporate melody to simulate music chord variation in different ways. Reinforcement learning (RL) is explored to model the music melody through a well-designed reward signal. For example, RL Tuner incorporates music melody through a deep Q-network with a hand-craft reward function that is designed based on the music theory and probability output of a trained Note RNN \cite{jaques2017tuning}. RE-RLTuner employs the Latent Dirichlet Allocation (LDA) topic model to provide LDA feature as a reward of reinforcement learning to learn the structural constraints of the music melody \cite{liu2021re}. However, RL reward design depends heavily on musical expert experience. Recurrent Neural Network (RNN) is always used as an agent of the reinforcement learning mode\cite{jaques2017tuning, liu2021re}. LookbackRNN introduces custom inputs and labels with 1 and 2 bars ago to recognize patterns that across 1 and 2 bars \cite{waite2016generating}. AttentionRNN accesses previous information with an attention mechanism to create the output for the current step \cite{waite2016generating}. In addition, the HRNN is introduced as a hierarchical LSTM architecture to learn the musical bar, beat, and note respectively \cite{wu2019hierarchical}. However, the above RNN based models only change inputs of LSTM without modifying the internal architecture of LSTM so that it hinders the ability to capture the correlation within a musical beat.  
	
	To fill the gap in that the previous RNN-based models can't capture the musical beat information, we propose a novel Enhanced Memory Network (EMN), which incorporates several Enhanced Memory Units (EMUs) to capture the musical beat information inside of the internal architecture of LSTM. This idea is motivated by the neuroscience theory that memory can incorporate melodic, harmonic, and modulatory regularities of tonal systems \cite{jan2011music}. In addition, overlearn and spaced learning can reduce the forgetting rate and improve retention, as well as step drop was seen on the forgetting curve \cite{rohrer2006effects, ebbinghaus2013memory, murre2015replication}. Unlike lookbackRNN which only changes the number of inputs, the EMN reconstructs LSTM input gate consisting of inputs with learnable parameters to incorporate a beat that consists of the four notes under the assumption of a 4/4 musical piece. Because the musical note distribution is dominated by the holding state \cite{ji2020comprehensive}, which negatively affects the diversity of the generated notes, cosine distance is also designed as part of loss functions to prevent EMN from remembering the holding state instead of musical notes to enrich the diversity of musical notes.
	
	\section{Methodology}
	\subsection{Enhanced memory network}
	Let $X=\{x_t\}_{t=1}^T$ be input musical encoding where $T$ is the length of music and $t$ is the index of a musical note. The EMN consists of several EMUs, output gates, forget gates, hidden states, and cell states which is shown in Figure \ref{fig:architecture}.
	
	The EMU is computed based on the beat that is summation of current and three previous note encodings as well as the current note encoding It has four input notes plus the previous hidden state and can be formulated as follows
	\begin{equation}
		e_t=\sigma (\sum_{j=t-3}^{t} W_j x_j + U_i h_{t-1})
	\end{equation} 
	where $W_j$ and $U_i$ are parameters to be trained and $\sigma(\cdot)$ is the sigmoid function. Then, the forget gate determines which contents need to forget as 
	\begin{equation}
		f_t = \sigma(W_f x_t + U_f h_{t-1})
	\end{equation}
	where $W_f$ and $U_f$ are learnable parameters. 
	Similar to $f_t$, the output gate $o_t$ is also computed based on $h_{t-1}$ and $x_t$ as 
	\begin{equation}
		o_t = \sigma (W_o x_t + U_o h_{t-1})
	\end{equation}
	where $W_o$ and $U_o$ are trainable parameters. The cell state $c_t$ decides how much need to update each state value as
	\begin{equation}
		c_t = f_t \odot c_{t-1} + e_t \odot \mathrm{tahn} (W_c x_t + U_c h_{t-1})
	\end{equation}
	where $W_c$ and $U_c$ are learnable parameters, $\odot$ denotes the element-wise multiplication. The update gate is decided according to a hidden state $h_t$ is updated with a the output gate $o_t$ and cell state $c_t$ as 
	\begin{equation}
		h_t = o_t \odot c_t
	\end{equation}
	
	\subsection{Loss function}
	The loss consists of classification loss and cosine distance loss between the current hidden state $h_t$ and the previous hidden state $h_{t-1}$. The classification loss $L_e$ is designed by adopting the multi-class cross-entropy loss as 
	\begin{equation}
		\mathcal{L}_e = - \sum_{k=0}^{K} x_{t+1,k}\log (p_{t,k})
	\end{equation}
	where $K$ is number of musical notes plus holding state and $x_{t+1,k}$ is ground truth labels that are the following note. Second, the cosine loss is employed between two neighboring hidden states to increase the note diversity and prevent the holding state dominates the learning music because of unbalanced data between the holding state and notes. The cosine distance loss is designed as the cosine distance between $h_{t-1}$ and $h_t$ as
	\begin{equation}
		\mathcal{L}_d = 1 - \frac{h_{t-1}\odot h_t}{||h_{t-1}||_2 ||h_t||_2}
	\end{equation}
	where $||\cdot||_2$ represents $\ell_2$ norm. Then the total loss can be represented as follows
	\begin{equation}
		\mathcal{L} = \lambda \mathcal{L}_e + (1-\lambda) \mathcal{L}_d
	\end{equation}
	where $\lambda$ is a hyperparameter to tune the importance of two loss functions and the influence will be discussed in the experiments.
	
	\section{Experiments}
	\subsection{Experimental Setup}
	\subsubsection{Dataset}
	The proposed model was trained on the melody folders in the Nottingham Music Database \footnote{dataset is available on https://github.com/jukedeck/nottingham-dataset}. The Nottingham Music Database has 1034 pieces of folk songs, pitch range from 55 to 88 in Midi format. We also quantified the compositions into 4/4 beat because most of the music compositions satisfy this style. Following Magenta (add footnotes or citation), we split the database into 90\% of music as training and 10\% as evaluation and encode the note continuation in the pitch sequences. In the experiment, we represented the music in the symbol domain that position 0 represented the holding state and the rests represented the position of notes.
	
	\subsubsection{Model Specification}
	Our model is implemented using TensorFlow \cite{abadi2016tensorflow}. Only one layer of LSTM is used and the hidden state dimension is 70. Adam optimizer is set as 0.005 initial learning rate and also the $\beta_1$, $\beta_2$ and $\epsilon$ are set to 0.9, 0.999, and 1$e^{-8}$. We train the model for 2800 iterations. During training, the batch size is 64. The trade-off hyperparameter $\lambda$ is set 0.5 and its influence will be discussed in the experiments. Using Magenta \footnote{https://github.com/magenta/magenta}, we generate 525 pieces of music ranging from note 0 to 34, and each note randomly repeats 15 times with 70 lengths of notes for each piece of generated music. 
	
	\subsection{Objective evaluation}
	In the objective evaluation, we compare the performance of the EMN with NoteRNN \cite{jaques2017tuning}, reinforcement learning-based models RL-Tuner \cite{jaques2017tuning} and RE-RLTuner \cite{liu2021re}. Following \cite{jaques2017tuning, dong2018musegan}, pitch count per bar (PC/bar), pitch count per beat (PC/beat), and mean autocorrelation lag 1 to 3 are calculated to evaluate the properties of the melodies generated by the above models.
	
	\begin{table*}[!h]
		\centering
		\label{tab:objectiveresult}
		\caption{Objective evaluation results. Average PC/bar, PC/beat, and Auto-lag1 to Auto-lag3 are reported in the test dataset. For current state-of-the-art models and our EMN, the performance which is closer to the test dataset is better.}
		\begin{tabular}{lcccccc}
			\toprule
			& PC/bar & PC/beat    & Auto-lag1 & Auto-lag2 & Auto-lag3 \\ \hline
			Dataset      & 3.34   & 2.34      & -0.48     & 0.44      & -0.49     \\ \hline
			NoteRNN \cite{jaques2017tuning}  & 4.41   & 2.92      & {\bf -0.21}     & 0.24      & -0.22     \\
			RL-Tuner \cite{jaques2017tuning}     & 6.98   & 3.71     & 0.01      & -0.01    & -0.01     \\
			RE-RLTuner \cite{liu2021re}  & 7.20   & 3.87   & -0.02     & -0.02     & -0.02     \\ \hline
			{\bf EMN(Ours) }    & {\bf 4.12}   & {\bf 2.77}    & -0.31     & {\bf 0.53}      & {\bf -0.32}     \\ \bottomrule
		\end{tabular}
	\end{table*}
	Table 1 showed that the structure performances of melodies generated by our EMN are significantly more like to the test dataset than those generated by the current state-of-the-art models except for Auto-lag1. For PC/bar and PC/beat, the EMN model that used the beat notes as input can learn the musical note correlation due to the intrinsic architecture of the model. In addition, the cosine loss might increase the diversity of notes. Thus, the distribution of notes from EMN has more similarity to the training dataset. For Auto-lag1, the NoteRNN model performed the best property, which might cause by the NoteRNN could learn more autocorrelation between the current note and the previous one. However, the performances of the EMN model were more dependable than the performance of others in Auto-lag2 and Auto-lag3, which implied that the EMN model described relationships among musical beats explicitly, that can capture the autocorrelation between current and two or three previous notes. In addition, the performances of reinforcement-based models were also worse than RNN based methods because the reward function that highly depends on a list of musical rules is challenging to design.
	
	\begin{figure}[!h]
		\centering
		\includegraphics[width=8cm]{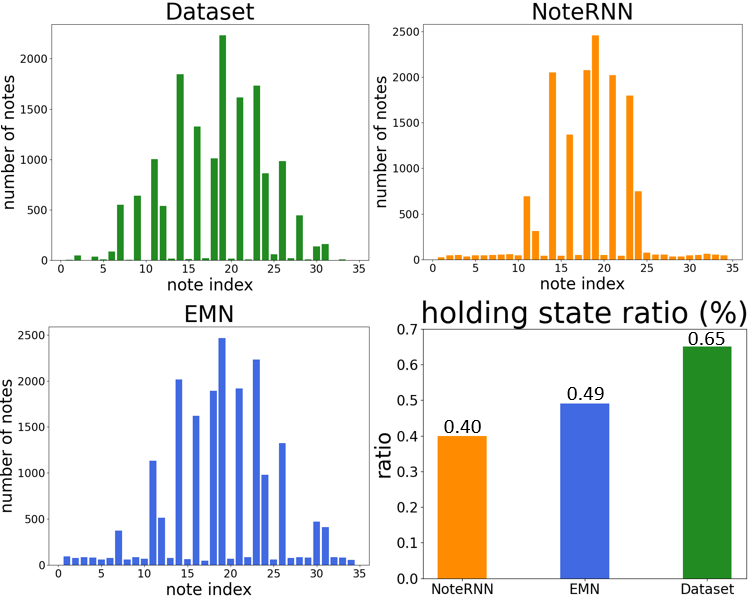}
		\caption{Distribution of Musical note and holding state among NoteRNN, EMN, and the test dataset.}
		\label{fig:note_dist}
	\end{figure}
	To illustrate the learned note and the holding state distribution of EMN, we plotted the note distribution of the test dataset, EMN model as well as NoteRNN in Figure \ref{fig:note_dist}. The musical note distribution of melodies generated from EMN is more similar to the test dataset than those generated from NoteRNN. This result showed that the EMN model can learn the musical correlation between the notes in each beat, making the structure of notes is more in line with the music theory of the real melodies. In likewise, the proportion of the holding state of EMN was also significantly approximate to the test dataset with a 16.53\% difference for the EMN and 25.25\% difference for the NoteRNN. The more approximate distribution implied that the EMU intrinsic architecture could capture musical structure information of beats or bars in melodies.
	
	\begin{table}[!h]
		\label{tab:lambdt}
		\caption{The performance of the melodies generated from EMN with different loss trade-off hyperparameter $\lambda$ on PC/bar, PC/beat, and Auto-lag1. The results represent the difference between the test dataset and EMN with different $\lambda$ which is test dataset values minus the results with different hyperprameter setting.}
		\centering
		\begin{tabular}{lccc}
			\hline
			$\lambda$	& PC/bar & PC/beat & Auto-lag1 \\ \hline
			0.0   & +0.94  & +0.59   & +0.37     \\
			0.1 & +0.72  & +0.44   & +0.32     \\
			0.2 & +0.85  & +0.48   & +0.33     \\
			0.3 & +0.81  & +0.49   & +0.30     \\
			0.4 & +1.25  & +0.73   & +0.47     \\
			{\bf 0.5} & {\bf +0.69}  & {\bf +0.43}   & {\bf +0.17}     \\
			0.6 & +0.95  & +0.60   & +0.37     \\
			0.7 & +1.12  & +0.66   & +0.18     \\
			0.8 & +1.10  & +0.58   & +0.26     \\
			0.9 & +1.39  & +0.69   & +0.26     \\ \hline
		\end{tabular}
	\end{table}
	To evaluate the effect of trade-off hyperparameter $\lambda$, we also investigate the performances of the different models with the different values of $\lambda$. The results were illustrated in Table 2 and suggested that the optimal hyperparameter can improve the performance of the EMN model in the musical structure because the optimal cosine loss function can avoid neighboring hidden states $h_{t-1}$ and $h_t$ becoming a trend to learn the notes' diversity. However, when the $\lambda$ increased, the performance of objective evaluation didn't constantly increase because over-focusing on cosine distance loss harms the note classification. When the $\lambda=0.5$, the EMN model got the best performance due to balancing the classification loss and cosine distance loss.  
	
	\subsection{Subjective evaluation}
	\begin{figure}[!h]
		\centering
		\includegraphics[width=8.5cm]{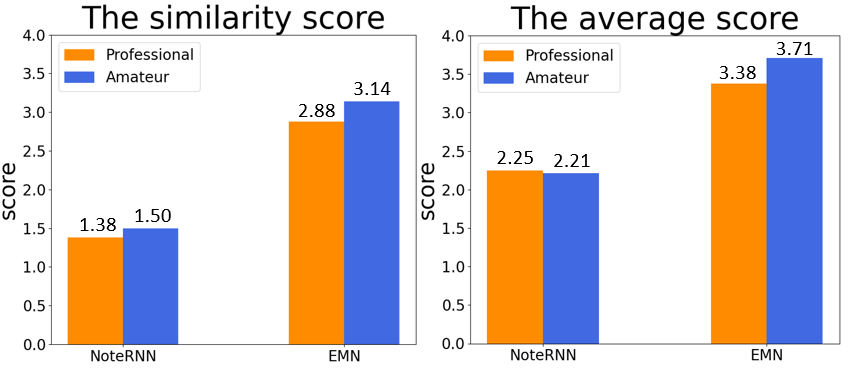}
		\caption{Subjective evaluation. The right figure is the average score for NoteRNN and EMN. The left figure is the similarity of style between the generated music and the dataset.}
		\label{fig:subjectiveimg}
	\end{figure}
In subjective evaluation, 22 volunteers, including 8 professional musicians and 14 music amateurs, evaluate the music quality of the melodies generated by the models. In addition, the style similarity between generated melodies and those in the training dataset was also estimated. Five pieces of music generated from each model were randomly selected as experiment materials. During the evaluation, the volunteers were asked to score the music quality of the generated music by NoteRNN and EMN using a five-point system and how close each of the two models compared with dataset music using 0-5 point. The grading results showed that our EMN achieved the highest score on both questionnaires for professional musicians and music amateurs. It implied that the melodies generated by EMN might be more melodious, enriched, and acceptable form in musical style than others.
	
	\section{Conclusion}
	In this paper, we propose a novel enhanced memory network with several enhanced memory units to capture the beat correlation inside the internal architecture of EMU. In addition, the cosine distance loss function is applied to enrich the note's diversity. The objective and subjective evaluation results show that the proposed EMN can be preferable to model the intrinsic structure of beats in musical melodies. To further enhance the ability of music generation, models will be constructed to better capture the duration of notes in the future work.
	
	\section{Acknowledgements}
	This work was supported in part by National Key R\&D Program of China (2020YFC2004100), in part by National Natural Science Foundation of China \\ (NSFC U1736202, NSFC 61771461) and Shenzhen KQTD Project (No. KQTD20200820113106007).

	
	\bibliographystyle{IEEEbib}
	\bibliography{strings,refs}
	
\end{document}